\documentclass[twocolumn]{pasj00}

\begin{document}
\SetRunningHead{Iguchi and Okuda}{The FFX Correlator}
\Received{2007/11/01}
\Accepted{2008/04/13}

\title{The FFX Correlator}


%
\author{Satoru \textsc{Iguchi} \altaffilmark{1}
       and 
       Takeshi \textsc{Okuda} \altaffilmark{1} 
       }
\altaffiltext{1}{National Astronomical Observatory, 2--21--1 Osawa, Mitaka, Tokyo 181--8588}
\email{s.iguchi@nao.ac.jp}
\email{takeshi.okuda@nao.ac.jp}


\maketitle

\begin{abstract}
We established a new algorithm for correlation process in radio astronomy. 
This scheme consists of the 1st-stage Fourier Transform as a filter 
and the 2nd-stage Fourier Transform for spectroscopy. 
The "FFX" correlator stands for Filter and FX architecture, 
since the 1st-stage Fourier Transform is performed as a digital filter, 
and the 2nd-stage Fourier Transform is performed as a conventional FX scheme. 
We developed the FFX correlator hardware not only for the verification 
of the FFX scheme algorithm but also for the application to 
the Atacama Submillimeter Telescope Experiment (ASTE) telescope 
toward high-dispersion and wideband radio observation 
at submillimeter wavelengths. 
In this paper, we present the principle of the FFX correlator 
and its properties, 
as well as the evaluation results with the production version. 

\end{abstract}

\section{Introduction}

The signals received by the antennas 
obey the stationary stochastic process and then 
ergodic process.  
The ergodic theory can be applied to the auto-correlation function 
for a spectrometer and the cross-correlation function 
for radio interferometer. 
Under such conditions, 
Weinreb (1963) developed the first digital spectrometer.
This digital spectrometer is called the XF correlator 
in which the correlation is calculated before Fourier Transform. 
Meanwhile, Chikada et al. (1987) 
developed the first the FX correlator of an another design, 
in which Fourier Transform is performed before cross multiplication.
Although there is a difference of property between
two basic designs, the obtained astronomical 
spectra of them 
were confirmed to be identical. 

Determining 
the number of correlation lags in the XF scheme 
or of Fourier Transform points in the FX scheme 
is essential for the realization of high-dispersion and wideband observation, 
because the frequency resolution 
is derived as 
\begin{equation}
\Delta f = 1/(\Delta t_\mathrm{s} N) = 2B/N,
\label{eq:it}
\end{equation}
where $\Delta t_\mathrm{s}$ is the sampling period, 
$N$ is the number of correlation lags or Fourier Transform points, 
and the bandwidth of B is equal to $1/(2 \Delta t_\mathrm{s})$.  
The material size and cost of the correlator strongly depend on 
the sampling period, $\Delta t_\mathrm{s}$, and 
the number of correlation lags or Fourier Transform points, $N$. 


The new XF architecture with the digital Tunable Filter Bank 
that is designed with the Finite Impulse Response (FIR) 
has been proposed and developed 
for the next generation radio interferometers, 
the Expanded Very Large Array (EVLA) 
and the Atacama Large Millimeter/submillimeter Array (ALMA) 
(\cite{EVLA}, \cite{ALMA1}). This is called the "FXF correlator". 
The architecture of the FXF scheme can make the material size smaller 
in comparison with that of the conventional XF scheme.
Since the digital filter allows a variety of observation modes 
[scientific and observational availability were shown in Iguchi et al. (2004)], 
the FXF scheme will provide us with the most appropriate specifications 
which meet the scientific requirements. 
This will lower the risk of over-engineering of the correlator.

The improved FX architecture with DFT filterbank 
was developed by Bunton (2000). 
The use of polyphase filter banks allows arbitrary filter responses 
to be implemented in the FX scheme (Bunton 2003). 
This is called the "Polyphase FX Correlator". 
This scheme has a possibility to achieve the spectral leakage of about -120 dB. 
In particular, this performance is significant to suppress 
the leakage from 
the spurious lines mixed in receiving, down-converting or digitizing.

The FFX Correlator is a new algorithm for correlation process 
in radio astronomy. 
The FFX scheme consists of 2-stage Fourier Transform blocks, 
which perform the 1st-stage Fourier Transform as a digital filter, 
and the 2nd-stage Fourier Transform to achieve higher dispersion. 
The first 'F' of the FFX is the initial letter of the word "Filter". 
In this paper, we present a new FFX architecture. 
The principle of the FFX scheme in section 2, 
the properties of the FFX scheme in section 3,
the algorithm verification and performance evaluation 
with the developed FFX correlator 
in sections 4 and 5, 
and the summary of this paper in section 6 are presented.

\section{Principle of the FFX Correlator}

\begin{figure*} Fig1
  \begin{center}
    \FigureFile(160mm,200mm){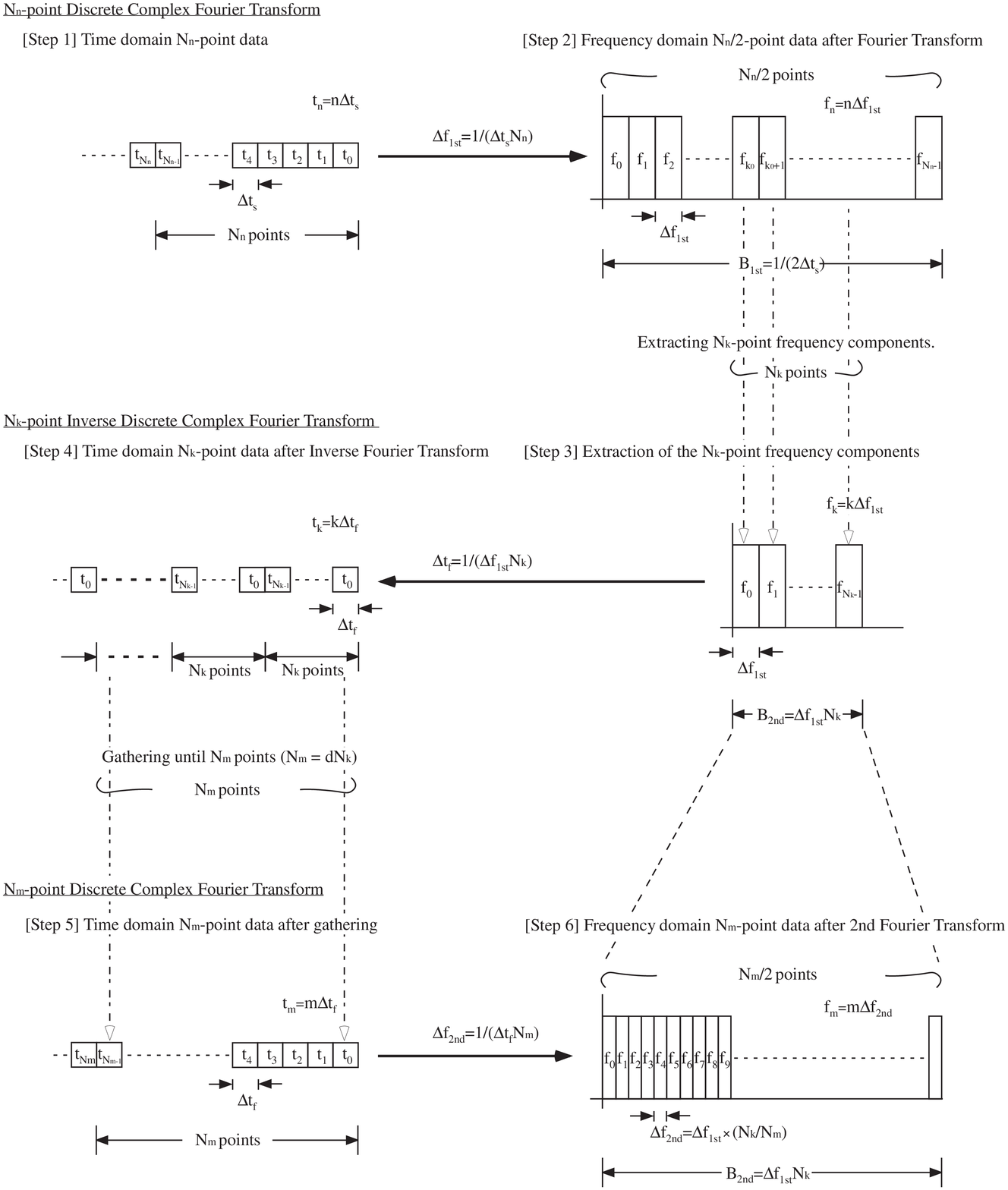}
  \end{center}
  \caption{Algorithm and the data flow diagram of the FFX scheme. }\label{fig:ffx}
\end{figure*}

This section shows the algorithm and the data flow diagram 
of the signal processing in the Fourier Transform of 
the FFX scheme (see figure \ref{fig:ffx}).
Suppose that $x_\mathrm{n}$ are the digital waveforms at the 
correlator input from the astronomical radio signals that 
are received by the telescope. 
The inputs, $x_\mathrm{n}$, are real digital 
signals at sampling period of $\Delta t_\mathrm{s}$, 
and obey the zero-mean Gaussian random variable. 
The suffix $n$ is an integer for time. 

[Step 1] The correlator receives the time-domain digital sampling signals 
from the Analog-to-Digital Converter (ADC), 
and accumulate them up to $N_\mathrm{n}$ points. 

[Step 2] The time-domain $N_\mathrm{n}$-point data 
are transferred to the frequency-domain by using
the $N_\mathrm{n}$-point Discrete Complex Fourier Transform as follows: 
\begin{equation}
X_\mathrm{p} = \Delta t_\mathrm{s} \sum^{N_\mathrm{n}-1}_{n=0} x_\mathrm{n} 
\exp\left(- j \frac{2 \pi p n}{N_\mathrm{n}}\right), 
\end{equation}
where $X$ is the spectrum after the 1st Fourier Transform, 
the suffix $p$ is an integer for frequency, 
and $\Delta t_\mathrm{s}$ is equal to $1/(2B_\mathrm{1st})$ at the bandwidth of $B_\mathrm{1st}$. 
The $\Delta f_\mathrm{1st}$ is the minimum frequency resolution of 
the 1st Fourier Transform, 
which is equal to $1/(\Delta t_\mathrm{s} N_\mathrm{n})$. 

[Step 3] The extraction of the $N_\mathrm{k}$ points 
from the frequency domain $N_\mathrm{n}/2$-point data 
after the 1st Fourier Transform is conducted 
as if filter and frequency conversion are performed simultaneously: 
\begin{equation}
X'_\mathrm{k} = X_{\mathrm{p}} \hspace{5mm} (p=k+k_0, \hspace{2mm} k = 0, \cdot \cdot \cdot, N_\mathrm{k}-1), 
\label{eq:df}
\end{equation}
where $k_0$ is the minimum frequency channel in the extraction, 
and the suffix $k$ is an integer for frequency. 

[Step 4] The $N_\mathrm{k}$-point data after Inverse Fourier Transform 
is written by 
\begin{equation}
x'_\mathrm{l} = \frac{1}{\Delta t_\mathrm{f} N_\mathrm{k}} \sum^{N_\mathrm{k}-1}_{k=0} X'_\mathrm{k} \exp\left[j \frac{2 \pi (k-N_\mathrm{k}/2) l}{N_\mathrm{k}}\right], 
\end{equation}
where $x'$ is the time-domain signal after inverse Fourier Transform, 
the suffix $l$ is an integer for time, 
and $\Delta t_\mathrm{f}$ is the sampling period after 
filtering at the bandwidth of $B_\mathrm{2nd}$ 
$(\Delta t_\mathrm{f}=1/B_\mathrm{2nd}=1/\Delta f_\mathrm{1st} N_\mathrm{k})$.
 
[Step 5] By repeating the procedure from Step 1 to Step 4, 
the data are gathered up to $N_\mathrm{m}$ points as follows; 
\begin{equation}
x'_\mathrm{m} = x'_\mathrm{l + d N_\mathrm{k}}
\end{equation}
where $m$ is $l+d N_\mathrm{k}$, and $d$ is 
the number of repeating times of the procedure from Step 1 to Step 4. 

[Step 6] The time-domain $N_\mathrm{m}$-point data after gathering 
are transferred to the frequency-domain by using 
the $N_\mathrm{m}$-point Discrete Complex Fourier Transform as follows: 
\begin{equation}
X'_\mathrm{q} = \Delta t_\mathrm{f} \sum^{N_\mathrm{m}-1}_{m=0} x'_\mathrm{m} 
\exp\left(- j \frac{2 \pi q m}{N_\mathrm{m}}\right), 
\end{equation}
where $X'$ is the spectrum after the 2nd Fourier Transform,
and the suffix $q$ is an integer for frequency. 
The $\Delta f_\mathrm{2nd}$ is the minimum frequency resolution 
after the 2nd Fourier Transform, 
which is equal to $1/(\Delta t_\mathrm{f} N_\mathrm{m})$ (=$\Delta f_\mathrm{1st} N_\mathrm{k}/N_\mathrm{m}$).

\begin{table}[t]
 \caption{Definition of functions. } 
\begin{center} 
\begin{tabular}{ll}
 \hline\hline
Mark  & Explanation    \\ 
\hline
$B_\mathrm{1st}$ & Bandwidth of the input signals \\
$\Delta t_\mathrm{s}$ & Sampling period of the input signals \\
$N_\mathrm{n}$ & Number of the points of 1st FT \\
$\Delta f_\mathrm{1st}$ & Minimum frequency resolution of 1st FT \\
$N_\mathrm{k}$ & Number of the extraction times as a filter \\
$B_\mathrm{2nd}$ & Bandwidth after extraction \\
$\Delta t_\mathrm{f}$ & Sampling period after filtering \\
$N_\mathrm{m}$ & Number of the points of 2nd FT \\
$\Delta f_\mathrm{2nd}$ & Minimum frequency resolution of 2nd FT \\
\hline
\end{tabular}
 \label{table:word}
\end{center}
Note that FT is Fourier Transform. 
\end{table}

\begin{table}[t]
 \caption{Relationship among the functions 
(see table \ref{table:word}). } 
\begin{center} 
\begin{tabular}{lll}
 \hline\hline
    & Equation  &  \\ 
\hline
(a) & $\Delta t_\mathrm{s}$     &= $1/(2B_\mathrm{1st}) $  \\
(b) & $\Delta f_\mathrm{1st}$ &= $2B_\mathrm{1st}/N_\mathrm{n} $  \\
    &                  &= $1/(\Delta t_\mathrm{s} N_\mathrm{n})$ \\
(c) & $\Delta t_\mathrm{f}$     &= $1/B_\mathrm{2nd}$  \\
    &                  &= $1/(\Delta f_\mathrm{1st} N_\mathrm{k}) $  \\
    &                  &= $\Delta t_\mathrm{s} N_\mathrm{n}/N_\mathrm{k} $  \\
(d) & $\Delta f_\mathrm{2nd}$ &= $B_\mathrm{2nd} / N_\mathrm{m}$  \\
    &                  &= $1/(\Delta t_\mathrm{f} N_\mathrm{m})$ \\
    &                  &= $\Delta f_\mathrm{1st} N_\mathrm{k}/N_\mathrm{m}$ \\
    &                  &= $2B_\mathrm{1st}/N_\mathrm{n} \cdot N_\mathrm{k}/N_\mathrm{m}$ \\
\hline
\end{tabular}
 \label{table:summary}
\end{center}
\end{table}

The definition of all functions used in this section is 
summarized in table \ref{table:word}. Also, 
the summary of the relationship among the functions 
(see table \ref{table:word}) 
is listed in table \ref{table:summary}. 
The following relations are derived: 
\begin{eqnarray}
\Delta f_\mathrm{1st} &=& 2B_\mathrm{1st}/N_\mathrm{n}, \\
\Delta f_\mathrm{2nd} &=& 2B_\mathrm{1st}/N_\mathrm{n} \cdot N_\mathrm{k}/N_\mathrm{m}. 
\end{eqnarray}
The frequency resolution of the FFX scheme 
is determined by the number of 
of the 1st Fourier Transform points($N_\mathrm{n}$), 
the number of the extractions as a filter ($N_\mathrm{k}$), and 
the number of the 2nd Fourier Transform points($N_\mathrm{m}$).

\section{Properties of the FFX Correlator}


The frequency responses for spectroscopic observations 
are finally derived by Fourier Transform in all schemes. 
For finite length of Fourier Transform, 
the responses are multiplied by a rectangular window function, 
which corresponds to convolving sinc function in the frequency domain.
The frequency profile of the XF scheme becomes
the shape of sinc function profile, while 
that of the FX scheme is sinc squared function profile. 
This indicates the FX scheme (including the FFX and polyphase FX schemes) 
is better than the XF scheme (including the FXF scheme) from the view points 
of the frequency profile and sharpness of individual frequency channels, 
and the spectral leakage.

For the realization of the high frequency resolution 
in the conventional FX scheme, 
the $N$-point complex FFT may be divided into $N/2$-point FFTs,
$N$-point twiddle factor multiplications, and $N/2$-point second FFTs, 
because the circuit size of the LSI (Large-Scale Integration) is limited  
(\cite{IG02}). 
The memory and circuit for the twiddle factor multiplications 
are critical. 
However, the high frequency resolution can be 
realized in the FFX scheme without the twiddle factor multiplications. 

The FFX scheme has advantages of selectivity for 
frequency resolution and bandwidth in comparison with other schemes. 
The comparable functions are also realized by implementing 
a digital LO circuits (\cite{ALMA1}). 
The digital LO circuits need to be delicately designed to avoid 
the spurious in mixing the digital LO signals due to 
the rounding errors in the calculation process.

\begin{figure*}[t]
\begin{center}
\FigureFile(147mm,100mm){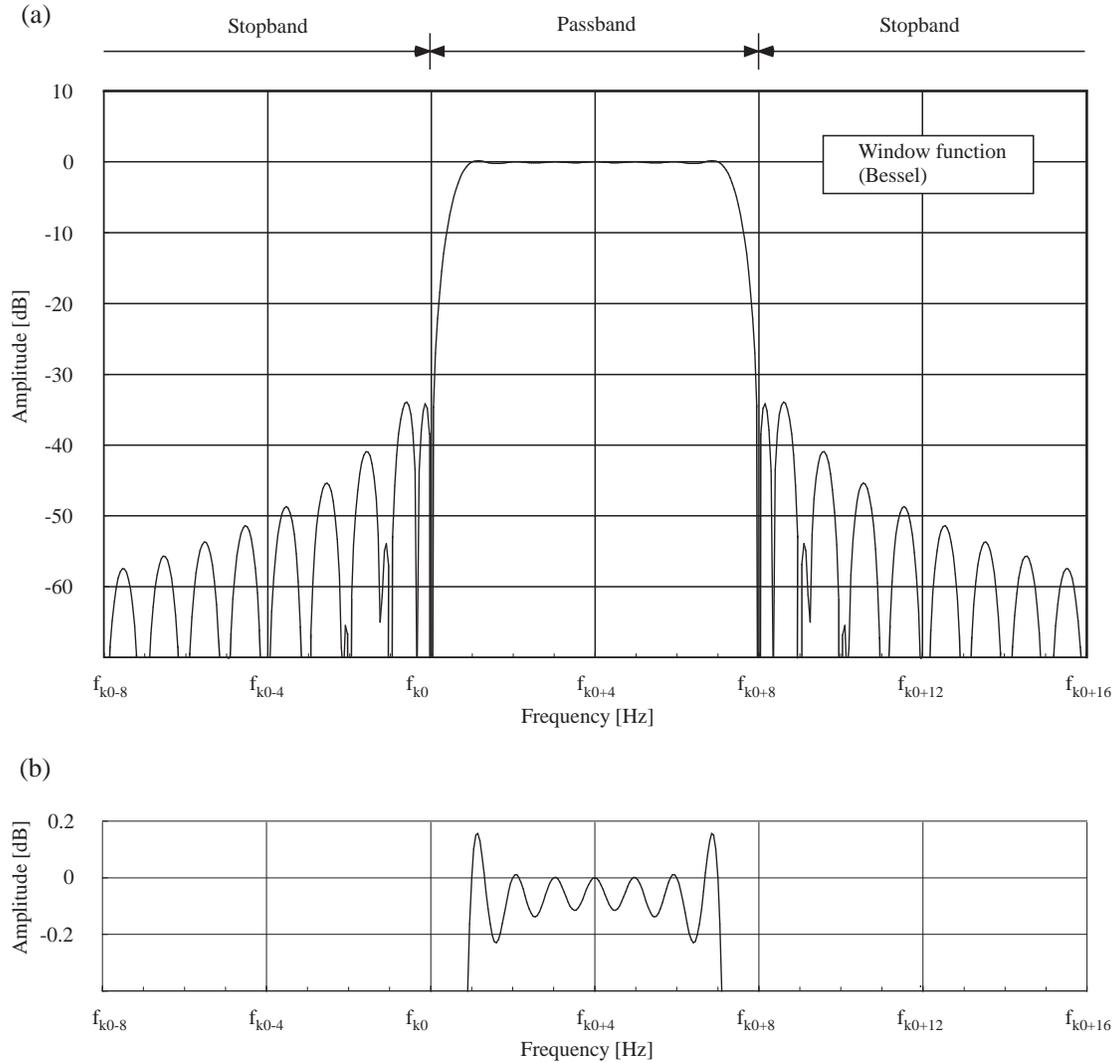}
\end{center}
\caption{
The frequency response of 1st FFT stage in the FFX scheme, which works as a digital filter. 
(a) is the frequency response at the full range to investigate the stopband, and 
(b) is the closeup frequency response in the passband to investigate a ripple. 
The frequency response is multiplied by the window of the Bessel function, $w(n)=J_0\left(\frac{4.24n}{N_\mathrm{n}}\right)$.
}
\label{fig:bessel}
\end{figure*}

For the investigation of aliasing or foldover 
from frequencies over the bandedge in the FFX scheme,
it is necessary and important to estimate the frequency response. 
As shown in [Step 3] in figure \ref{fig:ffx}, 
the desired frequency response, 
equation (\ref{eq:df}) can be rewritten as 
\begin{eqnarray}
X'_\mathrm{k} &=& 
X'_\mathrm{p} \hspace{2mm} (p=k+k_0, \hspace{2mm} k = 0, \cdots, N_\mathrm{k}-1), \\
X'_\mathrm{p} &=& H_\mathrm{p} \cdot X_\mathrm{p}, \\
H_\mathrm{p} &=& 
\left\{ 
\begin{array}{l@{}l}
1 & \hspace{2mm} (p = k_0, \cdots, k_0+N_\mathrm{k}-1) \\
0 & \hspace{2mm} (p < k_0, \hspace{2mm} p > k_0+N_\mathrm{k}-1). 
\end{array}  
\right. 
\label{eq:desiredFSF}
\end{eqnarray}
However, "$H_\mathrm{k_0}$" should be replaced by zero 
to reduce the aliasing or foldover of noise from frequencies over the bandedge. 
Thus, in the FFX scheme, equation (\ref{eq:desiredFSF}) need to be replaced as 
\begin{eqnarray}
H_\mathrm{p} &=& 
\left\{ 
\begin{array}{l@{}l}
1 & \hspace{2mm} (p = k_0+1, \cdots, k_0+N_\mathrm{k}-1) \\
0 & \hspace{2mm} (p < k_0+1, \hspace{2mm} p > k_0+N_\mathrm{k}-1). 
\end{array}  
\right.  
\label{eq:desiredFFX}
\end{eqnarray}
According to the two (1st and inverse) Fourier Transforms at the same resolution, 
the actual designed transfer function of the impulse response, which is derived with the square of sinc function, 
is represented as follows: 
\begin{eqnarray}
|H(f)| &=& \sum_{p=0}^{N_\mathrm{n}-1} H_\mathrm{p}^2 \left\{ \frac{\sin \left[\pi \left(N_\mathrm{n} \Delta t_\mathrm{s}f - p \right)\right]}{\pi \left(N_\mathrm{n} \Delta t_\mathrm{s} f - p \right)  } \right\}^2, \\ 
P(f) &=& \left\{ \sum_{p=0}^{N_\mathrm{n}-1} H_\mathrm{p}^2 \left\{ \frac{\sin \left[\pi \left(N_\mathrm{n} \Delta t_\mathrm{s} f - p \right)\right]}{\pi \left(N_\mathrm{n} \Delta t_\mathrm{s} f - p \right)  } \right\}^2 \right\}^2,
\label{eq:theory}
\end{eqnarray}
where $f$ is an arbitrary frequency, and also the response of a sinc function is caused by one Fourier Transform. 
It can be confirmed that 
the designed transfer function approaches 
the desired frequency response 
by increasing $N_\mathrm{n}$ while keeping $N_\mathrm{n}/N_\mathrm{k}$ constant. 
The filtering process in the FFX scheme 
is similar to that used in the design method of a Frequency Sampling Filter, FSF (\cite{FSF}). 

Also, to improve the frequency response, the window function before 1st Fourier Transform can be 
multiplied. In that case, equation (\ref{eq:theory}) should be written as 
\begin{eqnarray}
&&|H(f)| = \nonumber \\ 
&&\sum_{p=0}^{N_\mathrm{n}-1} H_\mathrm{p}^2 W(N_\mathrm{n} \Delta t_\mathrm{s} f - p) \frac{\sin \left[\pi \left(N_\mathrm{n} \Delta t_\mathrm{s} f - p \right)\right]}{\pi \left(N_\mathrm{n} \Delta t_\mathrm{s} f - p \right)  } , \\ 
&&P(f) = \nonumber \\ 
&&\left\{ \sum_{p=0}^{N_\mathrm{n}-1} H_\mathrm{p}^2 W(N_\mathrm{n} \Delta t_\mathrm{s} f - p) \frac{\sin \left[\pi \left(N_\mathrm{n} \Delta t_\mathrm{s} f - p \right)\right]}{\pi \left(N_\mathrm{n} \Delta t_\mathrm{s} f - p \right)  } \right\}^2, 
\label{eq:theory2}
\end{eqnarray}
where $W(f)$ is the response after 
the window function $w(n)$ is transferred to the frequency-domain by Fourier Transform. 
If the window is a rectangular window function, W(f) becomes a sinc function. 
In that case, equation (\ref{eq:theory2}) consists with equation (\ref{eq:theory}). 
There are the following famous window functions: Hanning, Hamming, Blackman, and Kaiser.
By well making a choice of the window function, the first sidelobe of stopband in the frequency 
response will be improved. It is well known that 
the first sidelobe levels with Backman and Kaiser as the window 
are better than those Hanning and Hamming. 
For the FFX scheme, it is found 
that the Bessel function of zeroth order $J_0$ is better than others. 
This frequency response is shown in figure \ref{fig:bessel}. 
The first and second sidelobe levels is about -34 dB, 
the fifth and sixth sidelobe levels is about -50 dB, 
and higher-order sidelobe levels will be better than -60 dB (see figure \ref{fig:bessel}a). 
The stopband response of the FFX scheme is not better than that of 
the polyphase FX scheme (Bunton 2000) and that of the FXF scheme for EVLA (\cite{EVLA}). 
On the other hand, the ripple response in the passband is 
less than 0.4 dB peak-to-peak (see figure \ref{fig:bessel}b). 
This performance is better than that of the polyphase FX scheme.

Note that the stopband performance to suppress the spurious lines can 
be improved by installing ``the detection and 
cancellation techniques of high and low frequency spurious lines'' into 
the FFX scheme (Chikada found this algorithm in the development of ALMA/ACA correlator).

\section{Requirements and Specifications for the Development of the FFX Correlator}

The development of the FFX correlator hardware is significant for the 
verification of the FFX scheme algorithm. 
It is necessary to define the requirements and specifications 
of the hardware of the FFX correlator. 
The hardware size can be optimized for the verification of algorithm. 
On the other hand, 
there were the scientific requests for the application 
to the Atacama Submillimeter Telescope Experiment (ASTE), which 
is a new project to install and operate a 10-m submillimeter 
telescope at a high latitude site (4,800 m) in the Atacama desert in 
northern Chile (\cite{ASTE}). 
Under these situations, the FFX correlator hardware 
for the algorithm verification 
was specified by also considering the scientific requirements 
for submillimeter astronomy including the application to ASTE.

In case of spectroscopic observations of atomic / molecular line emissions, 
their line width are extended by the Doppler shift along a line of 
sight with movement of inter stellar matter (ISM). 
In nuclear regions of external galaxies and Ultra Luminous Infrared 
Galaxies (ULIRGs), the molecular clouds which have various velocity 
components can be observed simultaneously, 
and line widths of the observed atomic / molecular emission lines are extended.  
For example, the line width of CO line emission of external galaxy 
is sometimes extended to more than 800 km s$^{-1}$ (i.e.~\cite{nar05}), 
which corresponds to about 920 MHz in $^{12}$CO(J=3--2) 
($\nu_{\rm rest}\sim 345.796$ GHz) and 
about 2.2 GHz in $^{12}$CO(J=7--6) ($\nu_{\rm rest}\sim 806.652$ GHz), 
by rotating around its nuclear region. 
On the other hand, in order to evaluate the kinematics of protoplanetary 
disks and the internal structure of molecular clouds in the Milky Way, 
it is also necessary to resolve their thermal line widths 
using spectrometer with high frequency dispersion; for instance, 
frequency resolution of 32 kHz corresponds to velocity resolution of 
0.032 km s$^{-1}$ at 1 mm wave length.

\begin{table}[t]
\caption{
FFT segment length.
}
\begin{center}
\begin{tabular}{ll}
\hline \hline
FFT stage & FFT segment length \\
\hline
First stage FFT   & 1024 points \\
Inverse FFT      & 8 points \\
Second stage FFT  & 4096 points \\ \hline
\end{tabular}
\end{center}
\label{table:NumFFT}
\end{table}

As the full bandwidth of more than 3 GHz is required, 
the FFX correlator need to achieve the processing speed of 
8192 Mega sample per second (Msps). 
In that case, 
for the realization of two-type spectral resolutions of 
about 5 MHz and less than 32 kHz, 
the FFX correlator must meet at least the specifications 
in table \ref{table:NumFFT}.

\begin{figure*}
  \begin{center}
    \FigureFile(170mm,70mm){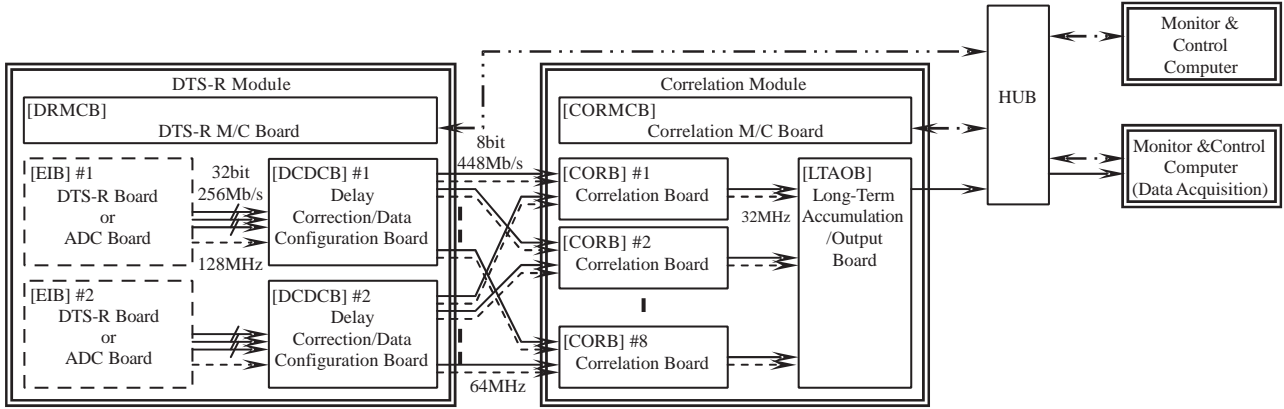}
  \end{center}
  \caption{Correlation processing block diagram of the FFX scheme.}
    \label{fig:blockdiagram}
\end{figure*}

\section{Development and Evaluation of the FFX Correlator}

The correlation processing block diagram of the FFX correlator 
is shown in figure \ref{fig:blockdiagram}. 
The FFX correlator consists of DTS-R (Data Transmission System Receiver) 
module, Correlation Module, and the Monitor $\&$ Control Computer. 
In the DTS-R module, the DTS-R Board or the 
ADC (Analog-to-Digital Converter) Board is implemented as 
an EIB (Electrical input Interface Board). 
The input data rate of the FFX correlator is 
about 48 Giga bit per second (Gbps)
with 3-bit quantization at the sampling frequency of 
8192 or 4096 Msps, 
which is 8192 Msps x 3 bits x 2 IF or 4096 Msps x 3 bits x 4 IFs. 
The DCDCB (Delay Correction and Data Configuration Board) 
effectively distributes the input signals to the next boards
for the parallel correlation processing. 
For the data processing at the throughput of 8192 Msps, 
the correlation is performed with 16 parallels, 
and both of the real and imaginary parts are used in FFT. 
The data is sent to each 16-parallel CORB (Correlation Board) per 
one-segment length. 
In the correlation mode of the FFX scheme, 
the data is sent per total segment length which 
is determined considering the signal process including the second stage of FFT. 
Final correlation output is obtained by adding 16-parallel correlation 
results. The correlation output is sent to 
the Monitor $\&$ Control Computer via LAN cable. 
Switching between the FX processing and the FFX processing 
is normally operated by setting the command 
into the Monitor $\&$ Control Computer.

\begin{figure}
  \begin{center}
    \FigureFile(85mm,100mm){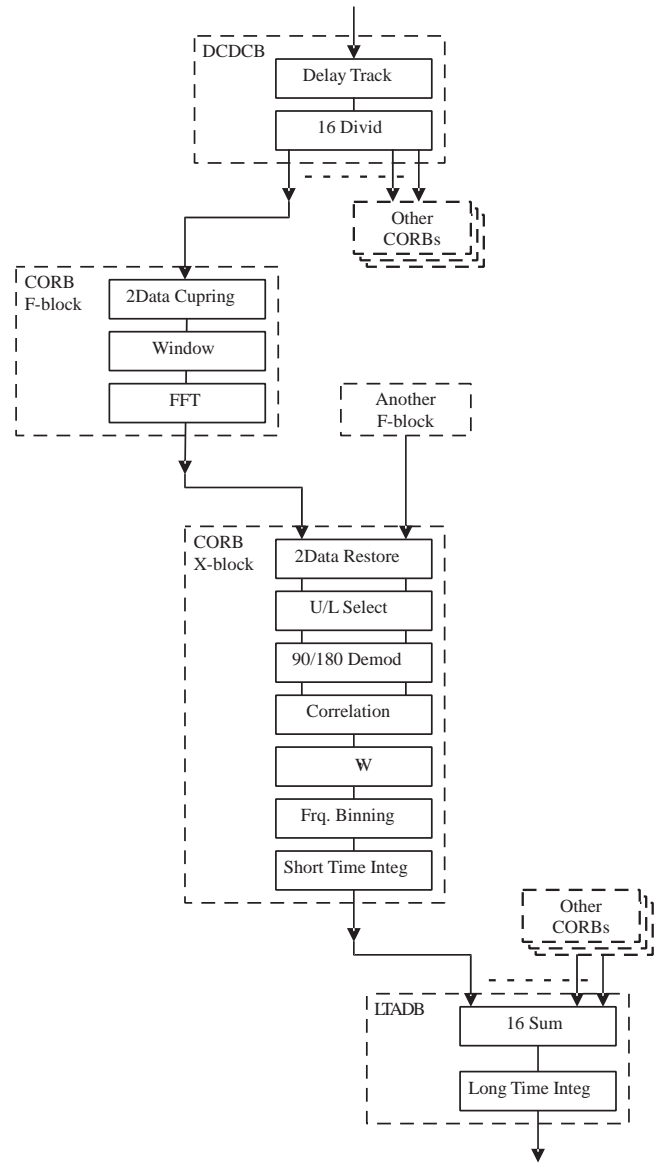}
  \end{center}
  \caption{Logical correlation processing flow.}
    \label{fig:corrflow}
\end{figure}

The correlation processing flow of the FFX correlator is shown 
in figure \ref{fig:corrflow}.
All main logics are implemented in FPGAs (Field Programmable Gate Array).

\subsection{Delay Correction and Data Configuration Board (DCDCB)}

\subsubsection{Delay Correction}

The FFX correlator has a delay correction circuit per bit 
for every 3-bit sampling signal.
Delay tracking for every single bit is realized by 
extracting 64-sample length from time-sequential 128 samples 
that are produced by splitting the output 
from FIFO (First In, First Out) memory in two, 
and shifting one side or the other in one-clock phase 
(see figure \ref{fig:DelayCorr}). 
Delay correction circuit has a FIFO memory of 1.024 Mega samples to each bit 
for delay tracking.
In case of 8192 Msps, 
the delay correction range is $\pm$ 512 kilo samples, 
that is $\pm$ 62.5 $\mu$sec
(=$\pm$ 18.75 km).

\begin{figure}[t]
  \begin{center}
    \FigureFile(75mm,80mm){fig5.eps}
  \end{center}
  \caption{Delay Correction.}
    \label{fig:DelayCorr}
\end{figure}

\subsubsection{Data Distribution}

\begin{figure}[t]
  \begin{center}
    \FigureFile(85mm,120mm){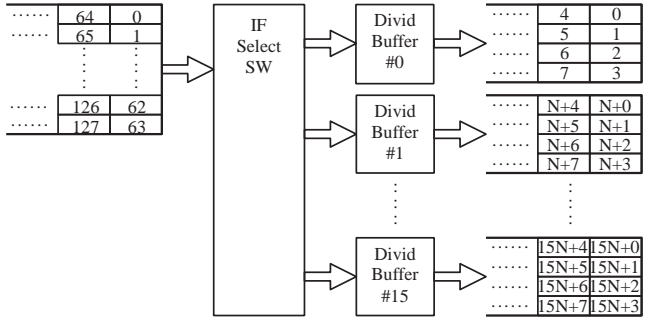}
  \end{center}
  \caption{Data Configuration.}
    \label{fig:DataDist}
\end{figure}

Given the operation speed of the device 
(FPGA etc.), 
16-parallel correlation processing is essential 
to achieve the throughput of 8192 Msps.
To do this, the input signals are divided by one segment length of FFT 
and sent to the Correlation Board (CORB) so that each parallel processing is performed separately. 

In the FFX processing, the data is divided by single segment length 
of the second-stage FFT. 
From table \ref{table:NumFFT}, the input signals are divided into:
\begin{equation}
1024 \times 4096/8 = 512 \hspace{2mm} \hat{k}, 
\end{equation}
where $\hat{k}$ is 1024 ($2^{10}$). 
This value determines the one segment length of this FFT processing.  
64-parallel signals are converted to 4-parallel 
after being output to each distribution buffer (see figure \ref{fig:DataDist}).

\subsection{Correlation Board (CORB)}

\subsubsection{Operation Format}

\begin{figure}[t]
  \begin{center}
    \FigureFile(85mm,120mm){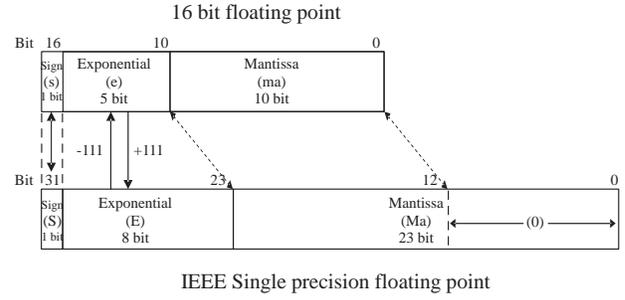}
  \end{center}
  \caption{Conversion between 16-bit floating point 
and IEEE single precision floating point.}
    \label{fig:bitformat}
\end{figure}

The operations of the correlation processing are performed 
using 16-bit floating point format.
FFT calculation is expressed as 
\begin{equation}
-1^s \cdot 2^{e\cdot 16} \cdot (1+ma/1024), 
\end{equation}
and the range is from $\pm$ 1/32768 to 131008. 
When the index ($e$) is 0, 
it is regarded as 0 instead of 2$^{-16}$ irrespective of mantissa.   

This not only simplifies the processing but also improves 
the compatibility with IEEE single precision floating point data.
IEEE single precision floating point has 32-bit representing sign ($S$): 
1 bit, exponential ($E$): 8 bits, and mantissa ($Ma$): 23 bits, 
and is expressed as 
\begin{equation}
  -1^S \cdot 2^{E\cdot 127} \cdot (1+M/223), 
\end{equation} 
and they have following relations 
\begin{eqnarray}
 S &=& s, \\
 E &=& e + 111, \\
 Ma &=& ma \times 8192 (2^{13}). 
\end{eqnarray}
Conversion between 16-bit floating point 
and IEEE single precision floating point is also shown 
in figure \ref{fig:bitformat}.

\subsubsection{Parallel processing of two datasets} 
\label{section:twodata}

 \begin{figure}[t]
  \begin{center}
    \FigureFile(85mm,120mm){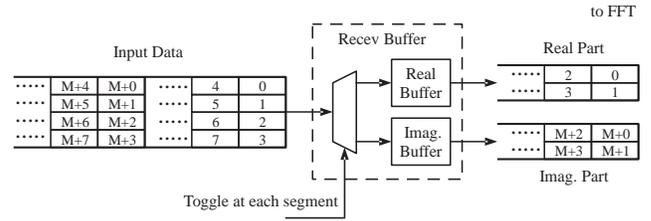}
  \end{center}
  \caption{Parallel processing of two datasets in FFT.}
    \label{fig:twodatasets}
\end{figure}

Since FFT is the linear response,
the real and imaginary parts of the input can be used 
separately. 
Since the signals received by an antenna are real part only, 
two datasets are combined into one complex dataset 
for the FFT processing (see figure \ref{fig:twodatasets}). 
One dataset is inserted into a real part, 
and then next dataset is inserted into an imaginary part. 
These two datasets are shifted with one-FFT segment length ($M$), 
and then processed as one complex data. 
Combination of two datasets is performed by Read/Write sequence control 
of the received buffer (=input buffer in FPGA).
This method can reduce the material size of the correlator.

\subsubsection{F-Block (FX mode)}

\begin{figure}[t]
  \begin{center}
    \FigureFile(85mm,120mm){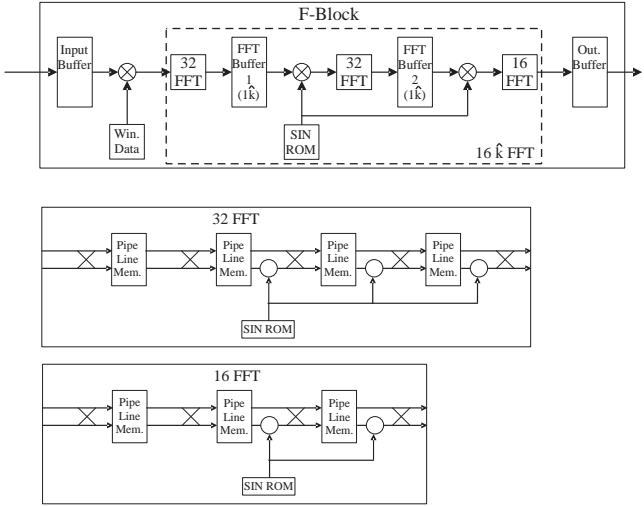}
  \end{center}
  \caption{Pipeline processing of the FX scheme (16$\hat{k}$-point FFT).}
    \label{fig:pipeline16kFFT}
\end{figure}

\begin{figure}[t]
  \begin{center}
    \FigureFile(85mm,120mm){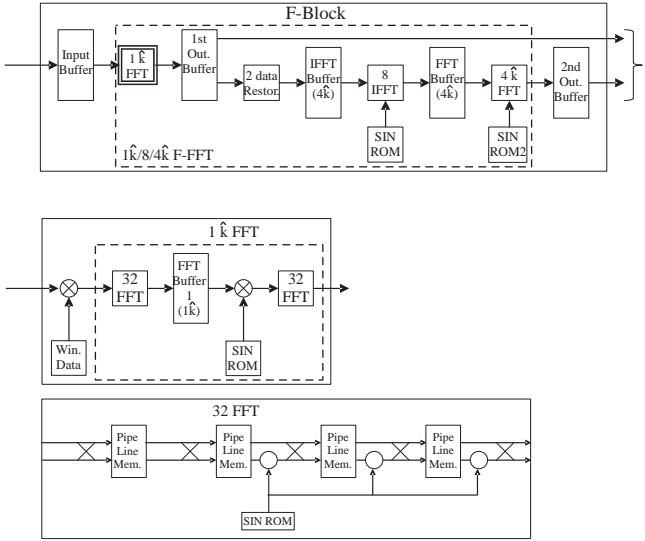}
  \end{center}
  \caption{Pipeline processing of the FFX scheme (1$\hat{k}$-point FFT, 8-point IFFT, 4$\hat{k}$-point FFT).}
    \label{fig:pipelineF-FX}
\end{figure}

The FX processing is performed with 16$\hat{k}$-point FFT.
In terms of the throughrate, pipeline processing 
is applied to each FFT stage processing.
To reduce the memory size in the pipeline processing, 
16$\hat{k}$-point FFT is realized by being divided 
into three parts: 32-point FFT, 32-point FFT, and 16-point FFT 
(see figure \ref{fig:pipeline16kFFT}). 
The input signals are multiplied by the window function 
between the input buffer and FFT, 
so that the input buffer size is based on the 3-bit signals. 
FFT is realized by changing the twiddle factor 
according to the processing stages. 
This helps minimize the ROM of the twiddle factors.

\subsubsection{F-Block (FFX mode)}

The FFX mode is shown in figure \ref{fig:pipelineF-FX}. 
The first-stage FFT (1$\hat{k}$-point FFT) is performed in the same manner 
as 32 points $\times$ 32 points of FX mode. 

The second-stage FFT (4$\hat{k}$-point FFT) is serially processed 
with one butterfly computing unit. 
Since 2$\times$8-point data is obtained every single process 
of 1k-point first-stage FFT, 4096-point data is obtained 
by repeating the process of the first-stage FFT 256 times.
The required time for this process is:  512 x 256 = 128$\hat{k}$ [CLK]. 
On the other hand, the required time for the process of 
4096-point FFT using one butterfly computing unit is: 
\begin{equation}
  4096 \times \log_2 (4096) / 2 = 24\hat{k} \hspace{2mm} \mathrm{[CLK]}. 
\end{equation}
Compared with these processing times, the processing time of 
4096-point FFT with one butterfly computing unit is much shorter.

\subsubsection{Window function processing}

After re-allocated to two datasets, 
the signals are multiplied by window function.
Any of the following window functions are selected: 
None (rectangular window), Hanning, Hamming, and Blackman.
Window functions are generated in the CPU according to the 
selected command ("WINDOW") from the Monitor $\&$ Control Computer 
before the correlation process starts.

\subsubsection{Data conversion}

 \begin{figure}[t]
  \begin{center}
    \FigureFile(30mm,40mm){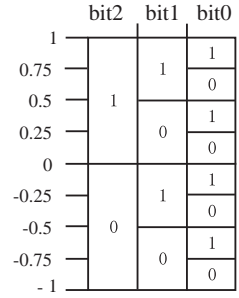}
  \end{center}
  \caption{Sampling bit level.}
    \label{fig:samplingbitlevel}
\end{figure}

 \begin{figure}[t]
  \begin{center}
    \FigureFile(80mm,120mm){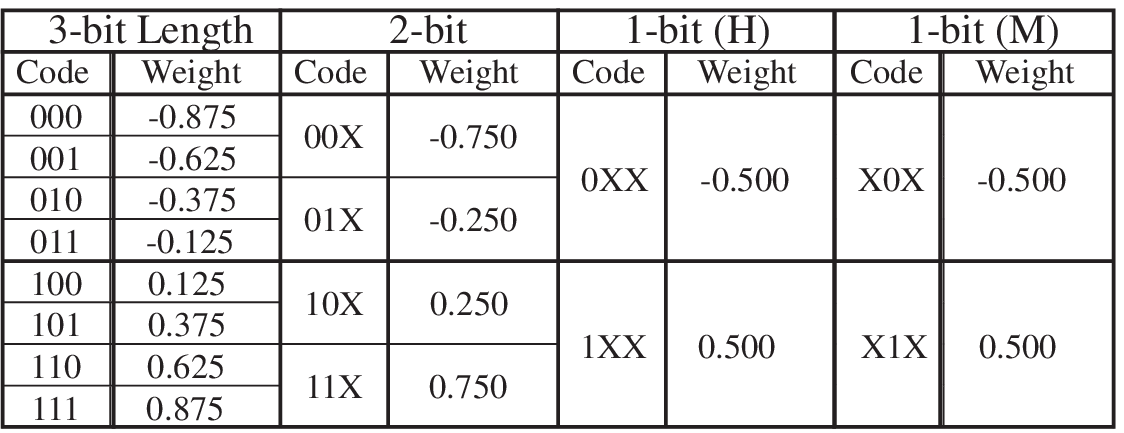}
  \end{center}
  \caption{Sampling bit level at 3-bit length.}
    \label{fig:samplingbitlevel2}
\end{figure}

The input signals go through 3-bit processing. By setting a 
command ("SMPLBIT") from the Monitor $\&$ Control Computer, 
the input signals are processed as 1-bit or 2-bit data. 
Also, conversion to 1-bit data with (M) middle bit only is available.  
Assumed threshold value for each bit of sampling data is given 
as shown in figure \ref{fig:samplingbitlevel}. 
Based on the above threshold value, converted value 
of the input data is calculated as shown in figure \ref{fig:samplingbitlevel2}. The input signals are converted into 16-bit floating point operation format before FFT.

\subsubsection{X-block (FX mode)}

 \begin{figure}[t]
  \begin{center}
    \FigureFile(80mm,120mm){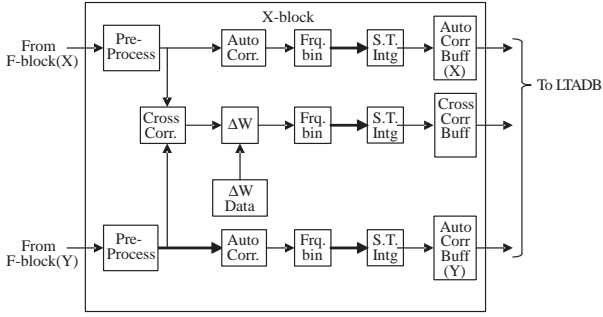}
  \end{center}
  \caption{Block diagram of the correlation in a FX mode.}
    \label{fig:xblockFX}
\end{figure}

 \begin{figure}[t]
  \begin{center}
    \FigureFile(80mm,120mm){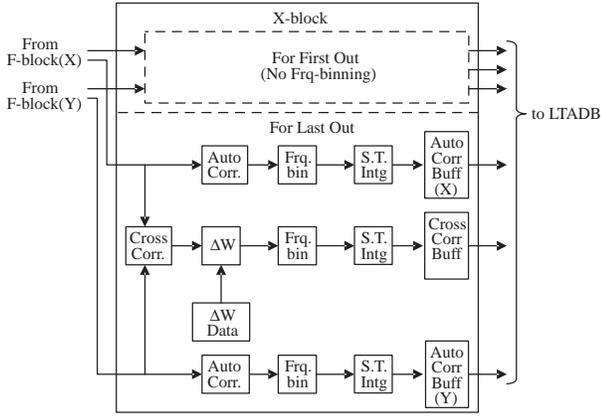}
  \end{center}
  \caption{Block diagram of the correlation in a FFX mode.}
    \label{fig:xblockFFX}
\end{figure}

The composition of X-block in normal FX processing is shown 
in figure \ref{fig:xblockFX}. 
Correlation processing is performed in 8$\hat{k}$-channel either of USB or LSB. 
For improving the sensitivity loss (see table 6 of \cite{oku01}), 
2-channel frequency binning is performed to generate 4$\hat{k}$-channel data 
(Frq. bin). 
Integration of 100 milliseconds is conducted in time integration (S.T.Ing).
If the integration takes longer than 100 milliseconds, 
it is performed in the Long-Term Accumulation and Output 
Board (LTAOB) at the time of 16-parallel data synthesis.

In OTF mode, the data is compressed from 8$\hat{k}$ 
frequency channels to 32 frequency channels by frequency binning. 
With time component of 1 millisecond, 
the correlation buffer area is changed every 1 millisecond.
The output to the Long-Term Accumulation and Output Board (LTAOB)
is performed every 100 millisecond.

Prior to the correlation processing, the data from the F-block needs 
some pre-processing such as split of two datasets, 
USB/LSB (upper and lower sideband) selection, 
and 90-degree and 180-degree phase switching demodulations.   
In splitting process of two datasets, 
the input data are added and subtracted, 
and then the two-data components combined in the F-block are split.  
When LSB is selected, the sign of the split Imag component is inversed. 
On the other hand, when USB is selected, the split data is sent to the correlation process 
without any additional processing.
In 90-degree and 180-degree phase-switching demodulations, 
switching between Real and Imag, and sign inversion 
are performed according to the two phase-switching signals.

\subsubsection{X-block (FFX mode)}

X-block in the FFX processing is shown in figure \ref{fig:xblockFFX}. 
In the FFX processing, X-block has a dual structure considering 
the processing speed of the device. 
The first-stage FFT is the same as that of FX mode, 
however frequency binning is not performed.  
Split of two datasets in the second-stage FFT is not performed in X-block, 
since the process is performed in F-block.
The results of the second stage FFT are output together with USB and LSB. 
Two-channel frequency binning is performed for improving 
the sensitivity loss (see table 6 of \cite{oku01}).

\subsubsection{$\Delta W$ correction}

 \begin{figure}[t]
  \begin{center}
    \FigureFile(80mm,120mm){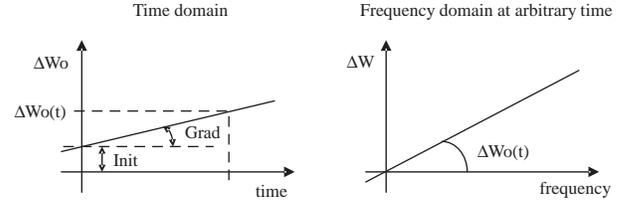}
  \end{center}
  \caption{$\Delta$ W correction.}
    \label{fig:deltaW}
\end{figure}

 \begin{figure}[t]
  \begin{center}
    \FigureFile(80mm,120mm){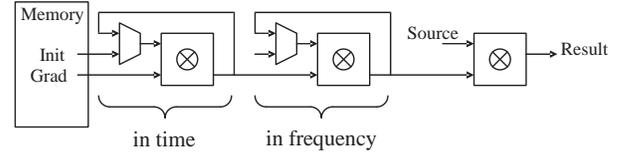}
  \end{center}
  \caption{$\Delta W$ correction circuit.}
    \label{fig:deltaWcir}
\end{figure}

\begin{table*}[t]
\caption{
Operation mode in a FX mode.
}
\begin{center}
    \begin{tabular}{lllll}
      \hline
      \hline
      Bandwidth & Spectral & Spectral   & Velocity  & Correlation\\
                & points   & resolution & resolution & \\
                &          &            & at 1 mm & \\
      \hline
      4096 MHz & 4096 & 1 MHz   & 1.0 km s$^{-1}$ & 2AC--1CC \\
      2048 MHz & 4096 & 0.5 MHz & 0.5 km s$^{-1}$ & 4AC--1CC \\
      \hline
    \end{tabular}
\end{center}
\label{table:FXopmode}
\end{table*}

\begin{table*}[t]
\caption{
Operation mode in a FFX mode.
}
\begin{center}
\begin{tabular}{llllll}
\hline
\hline
Stage & Bandwidth & Spectral & Spectral   & Velocity   & Correlation \\
      &           & points   & resolution & resolution & \\
      &           &          &            & at 1 mm    & \\
\hline
1$^{\rm st}$ & 4096 MHz & 512  &     8 MHz  & 8.0 km s$^{-1}$   & 2AC--1CC \\
2$^{\rm nd}$ & 64 MHz   & 2048 & 31.25 kHz  & 0.031 km s$^{-1}$ & 2AC--1CC \\
\\
1$^{\rm st}$ & 4096 MHz & 512  &     8 MHz  & 8.0 km s$^{-1}$   & 2AC--1CC \\
2$^{\rm nd}$ & 128 MHz  & 4096 & 31.25 kHz  & 0.031 km s$^{-1}$ & 1AC \\
\\
1$^{\rm st}$ & 2048 MHz & 512  &     4 MHz  & 4.0 km s$^{-1}$   & 4AC--2CC\\
2$^{\rm nd}$ & 32 MHz   & 2048 & 15.625 kHz & 0.016 km s$^{-1}$ & 4AC--2CC\\
\\
1$^{\rm st}$ & 2048 MHz & 512  &     4 MHz  & 4.0 km s$^{-1}$   & 2AC--1CC\\
2$^{\rm nd}$ & 64 MHz   & 4096 & 15.625 kHz & 0.016 km s$^{-1}$ & 2AC--1CC\\
\\
1$^{\rm st}$ & 2048 MHz & 512  &     4 MHz  & 4.0 km s$^{-1}$   & 2AC--1CC\\
2$^{\rm nd}$ & 128 MHz  & 8192 & 15.625 kHz & 0.016 km s$^{-1}$ & 1AC \\
\\
1$^{\rm st}$ & 2048 MHz & 512  &     4 MHz  & 4.0 km s$^{-1}$   & 2AC--1CC\\
2$^{\rm nd}$ & 96 MHz   & 6144 & 15.625 kHz & 0.016 km s$^{-1}$ & 1AC \\
2$^{\rm nd}$ & 32 MHz   & 2048 & 15.625 kHz & 0.016 km s$^{-1}$ & 1AC \\
\hline
\multicolumn{6}{l}{{\footnotesize 4AC: Auto-Correlation (H$_1$H$_1$ and V$_1$V$_1$, H$_2$H$_2$ and V$_2$V$_2$), 2CC: Cross-Correlation (H$_1$V$_1$, H$_2$V$_2$)} \hss} \\ 
\multicolumn{6}{l}{{\footnotesize 2AC: Auto-Correlation (HH and VV), 1CC: Cross-Correlation (HV)} \hss} \\ 
\multicolumn{6}{l}{{\footnotesize 1AC: Auto-Correlation (HH or VV).} \hss} \\
    \end{tabular} 
\end{center}
\label{table:FFXopmode}
\end{table*}

$\Delta W$ correction is performed based on the baseline.
$\Delta W$ correction coefficient is generated within FPGA. 
Firstly, $\Delta W_0(t)$, the gradient of $\Delta W$ to time change, 
is evaluated and then the value of each frequency channel is calculated 
(see figure \ref{fig:deltaW}). 
Circuit diagram of the line graphs above is shown in figure \ref{fig:deltaWcir}.

For every 1 millisecond, the initial value (Init) is provided 
by the Monitor $\&$ Control Computer, and read as a initial gradient 
[$\Delta W_0(0)$]. 
In the border between segments, 
the previous value is multiplied by the gradient (Grad) data 
to generate a gradient of a new segment [$\Delta W_0(t)$].  
In the arbitrary time $(t)$, initial value (the value of DC) is set to 0.
$\Delta W$ is set for every 128 channel in the full bandwidth of 8$\hat{k}$ channel, 
which means the full bandwidth (8$\hat{k}$ channel) is corrected with 64 steps.  
The initial value of $\Delta W_0(t)$ can be set for every 1 millisecond, however, a given value is set within 128 channel. 
The gradient of $\Delta W_0(t)$ is the gradient variation of $\Delta W$ per one segment length for FFT.
The Monitor $\&$ Control Computer specifies a set of the initial value 
and gradient approximately every 1 second.

\subsection{Long-Term Accumulation/Output Board (LTAOB)}

 \begin{figure}[t]
  \begin{center}
    \FigureFile(80mm,120mm){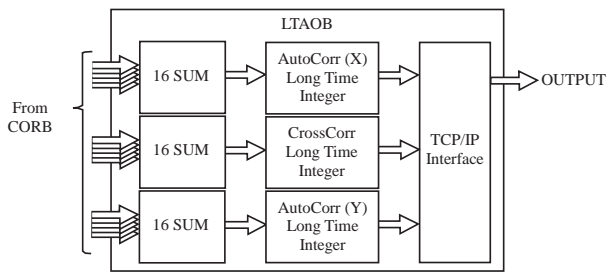}
  \end{center}
  \caption{Long-Term Accumulation/Output Board.}
    \label{fig:LTA}
\end{figure}

 \begin{figure*}[t]
  \begin{center}
    \FigureFile(130mm,120mm){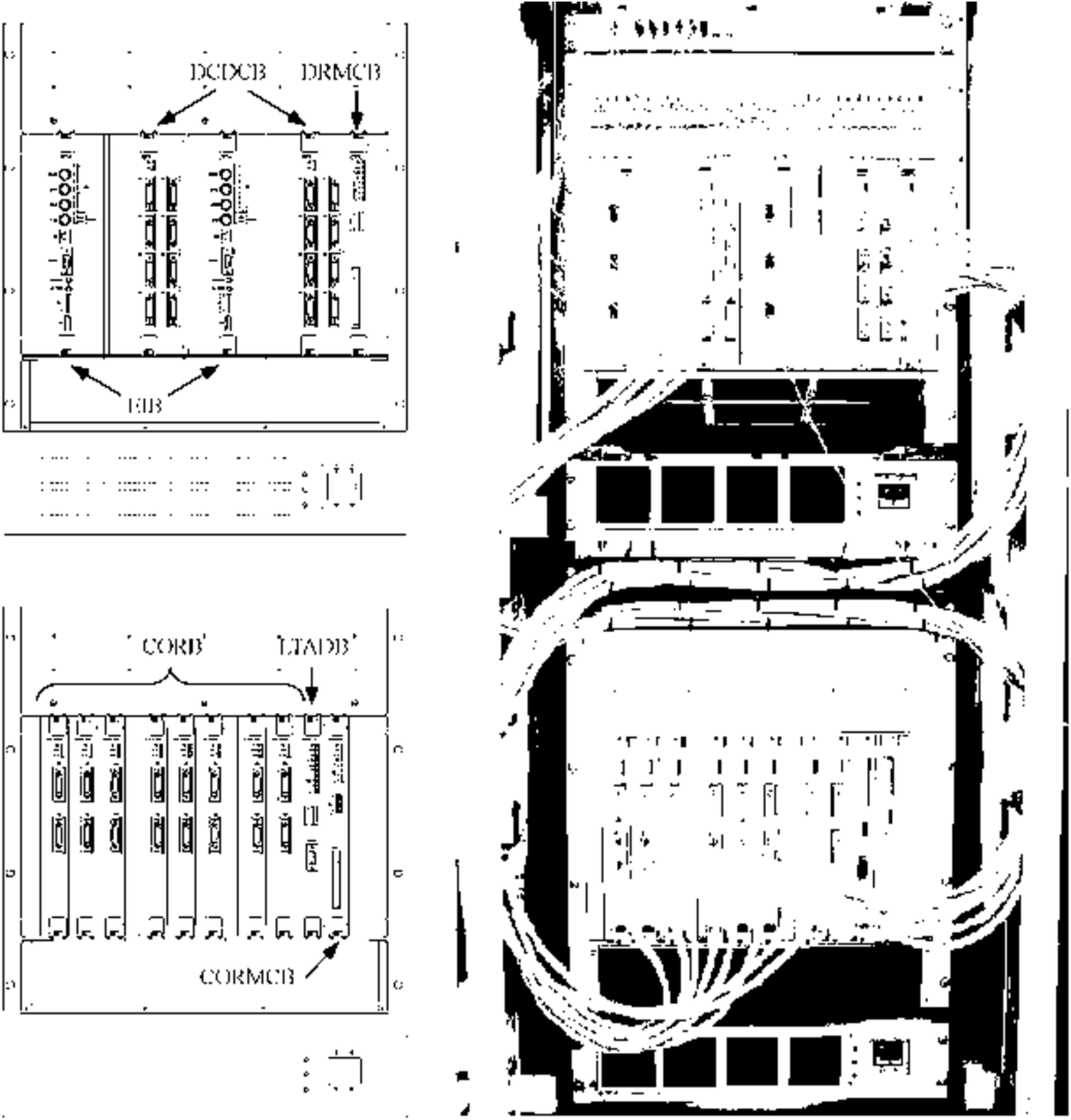}
  \end{center}
  \caption{FX correlator under development. 
There are two modules: one is the DTS-R module 
(top: left is a drawing of the front view, and right is a photograph), 
having a size of (W) 490 mm $\times$ (H) 500mm $\times$ (D) 500 mm; 
the other one is the Correlation module 
(bottom: left is a drawing of the front view, and right is a photograph), 
having a size of (W) 490 mm $\times$ (H) 500mm $\times$ (D) 500 mm. 
The total size of the GDFB is (W) 490 mm $\times$ (H) 1100mm $\times$ 
(D) 500 mm, including the two power supply modules. 
  }
    \label{fig:farm}
\end{figure*}

Final correlation value is calculated by adding the output of 16 correlators.
The composition of the Long-Term Accumulation/Output Board (LTAOB)
is shown in figure \ref{fig:LTA}. 
When the correlation result is output, 
the data is converted into the format of 
IEEE single-precision floating point.

In the FX processing, the number of the output frequency channels 
is normally 4 $\hat{k}$ (=4096). 
Thus the output data size per one correlation is    
\begin{equation}
4 \hat{k} \times 32 
\times 2  \times 
3 = 768 \hat{k} \hspace{1mm} \mathrm{bits},  
\end{equation}
where $32$ is the number of single-precision floating bits, 
$2$ is the complex, and $3$ is the number of correlations; 
auto-correlations of $x$ and $y$, and a cross-correlation between them. 
Assuming the minimum integration time is 0.1 second, 
the estimated output speed is approximately 7.7 Mbps. 
In the FFX processing, the number of the output frequency channels 
is 512 + 2 $\hat{k}$, thus the data size per one correlation is 480 $\hat{k}$ bits. 
Consequently, the estimated output speed with 0.1-second integration time 
is approximately 4.8 Mbps.
Correlation results are sent to the Monitor $\&$ Control Computer 
using TCP/IP protocol of 100 BaseT-Ether.

\subsection{Operation mode}

The FFX correlator are divided into four F parts in principle 
(not physically). Each F part 
in a FX mode and the first FFT stage of FFX mode are operated 
at 2048 MHz, and each F part at the second FFT stage of FFX mode 
are operated at 32 MHz. Also, the digital signals input from EIB
are distributed to arbitrary F parts by using the command 
(``FCHSEL'') from the Monitor $\&$ Control Computer. 
In FX mode and the first FFT stage of FFX mode, 
the digital signals of 8192 Msps are operated by combining 
two F parts (= 2$\times$2048 MHz), while the signals of 4096 Msps 
are operated by using one parts (=1$\times$2048 MHz). 
At the second FFT stage of FFX mode, the digital signals 
of 8192 Msps are operated by combining two F parts 
(=2$\times$32 MHz), while the signals of 4096 Msps 
are operated using only one parts (=1$\times$32 MHz). 
By using the command ("FCHSEL"), the effective bandwidth after the second FFT stage can be changed. 

All of the operation modes available in this FFX correlator are listed 
in table \ref{table:FXopmode} and \ref{table:FFXopmode}.

\subsection{Hardware of the FX Correlator}

The Hardware of the FX Correlator is shown in figure \ref{fig:farm}.
The DTS-R module consists of 
two Electrical input Interface Boards (EIBs), 
two Delay Correction and Data Configuration Boards (DCDCBs), and 
one DTS-R Monitor $\&$ Control Boards (DRMCBs). 
The Correlation module consists of 
eight Correlation Boards (CORBs), 
Long-Term Accumulation/Output Board (LTAOB), and 
one Correlation Monitor $\&$ Control Board(CORMCB). 
Each module is connected to the independent Power module. 

The power consumption of the filter module is 400 W, 
while that of the output module is 600 W. 
The total AC power is 750 W at 1-phase 100-220 VAC 50/60 Hz 
(100 V $\pm 10\%$ or 220 V $\pm 10\%$).
The total weight is 71.3 kg.

\subsection{Evaluation and Discussion}

\begin{figure}[t]
\begin{center}
\FigureFile(80mm,65mm){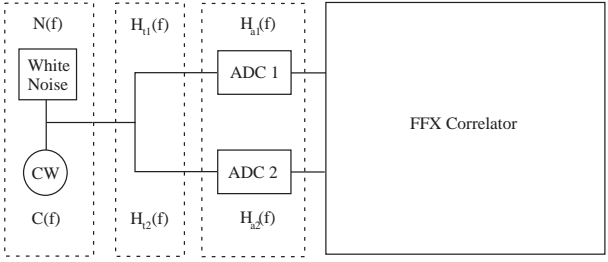}
\end{center}
\caption{Block diagram of the measurement setup. 
A detailed block diagram of the FFX correlator 
is shown in figure \ref{fig:blockdiagram}. 
The white noise is generated by the ASTE analog backend subsystem,
while the CW signal is produced by 
the Agilent E4440B, 
which is locked in phase to 
a standard signal of 10 MHz generated by 
the reference clock generator. 
The ADC 1 and ADC 2 with 8192 Msps are also locked in phase to 
a signal of 8192 MHz generated by 
the reference clock generator. 
Note that the ADCs were developed specifically for ASTE 
(\cite{oku07}), and all output signals 
from the reference clock generator are 
generated by the built-in 5MHz crystal oscillator (OCXO).
}
\label{fig:measure}
\end{figure}

Figure \ref{fig:measure} shows a block diagram of the measurement setup of 
the frequency response of the FFX Correlator. 
To investigate the frequency response, 
it is useful to use the CW signal, which can measure the folding effects 
by sweeping the frequency range of 0 to 4096 MHz. 
The white noise is important 
to measure the frequency response,  
because the astronomical signals obey the Gaussian random variable. 
To obtain the input signals that are approximated to the zero-mean Gaussian probability, mixing of the CW signal with the white noise is necessary.

\begin{figure*}[t]
\begin{center}
\FigureFile(147mm,100mm){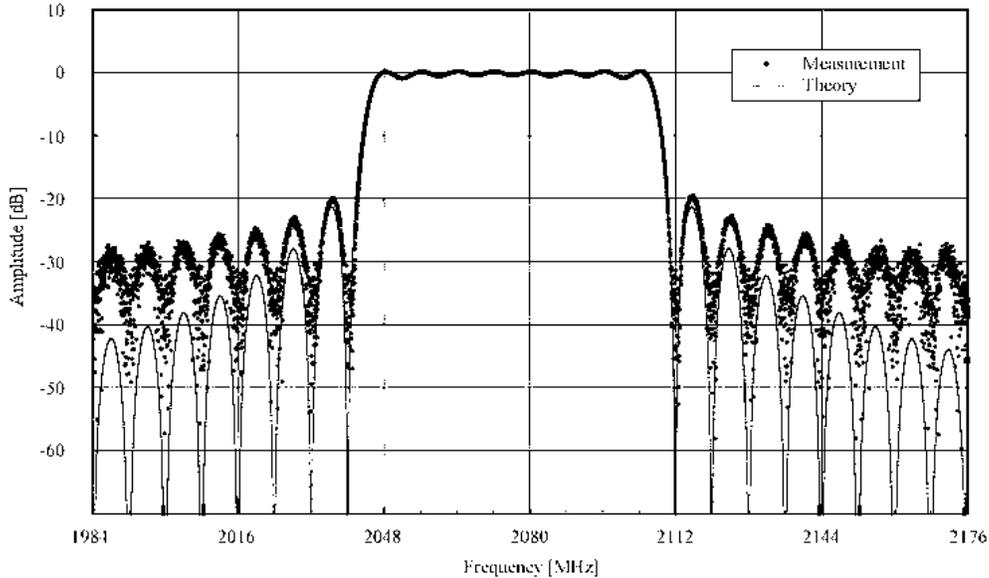}
\end{center}
\caption{
The measured frequency response of the FFX correlator at a bandwidth of 64 MHz 
in a frequency rage of 2048 to 2112 MHz. 
First sidelobe of about $-20$ dB and  
other stopband response of better than $-23$ dB were successfully achieved. 
A comparison between 
the measured frequency response of the FFX correlator (points) 
and the ideal frequency response of the FFX scheme (full line) 
in a rectangular window function without zeroing 
[see equations (\ref{eq:theory}) and (\ref{eq:desiredFSF})] are shown. 
}
\label{fig:response}
\end{figure*}

The frequency response of the FFX correlator when CW is included 
is written as
\begin{eqnarray}
P_{\mathrm{on}}(f) &=& a_{\mathrm{on}} \cdot \left[|C(f)|^2 +|N(f)|^2 \right] \nonumber \\
&&\cdot H_{1}(f) H_{2}^{\ast}(f) \cdot |H_{\mathrm{D}}(f)|^2, 
\label{eq:oncor}
\end{eqnarray}
where $C(f)$ is the frequency response of the CW signal, 
$N(f)$ is the frequency response of the white noise 
from the ASTE analog backend subsystem, 
$H_{1}$ and $H_{2}$ are 
the frequency response by different transmission paths, 
in which $H_1 =H_{\mathrm{t}1} H_{\mathrm{a}1}$ and $H_2 =H_{\mathrm{t}2} H_{\mathrm{a}2}$
(see figure \ref{fig:measure}), 
and $H_{\mathrm{D}}$ is the frequency response of the FFX correlator, 
including the effects of requantization and the folding noise after downsampling. 
The bandpass calibration is essential for estimating the CW power accurately, 
because the bandpass response becomes a time-variable due to outdoor air temperature. 
The frequency response without the CW signal is 
written as 
\begin{equation}
P_{\mathrm{off}}(f) = a_{\mathrm{off}} \cdot |N(f)|^2 \cdot H_{1}(f) H_{2}^{\ast}(f) \cdot |H_{\mathrm{D}}(f)|^2. 
\label{eq:coroff}
\end{equation}
The ADCs work as 1-bit performance (\cite{oku07}). 
In that case, it is important to adjust the power 
as precisely as possible so as to avoid the high-order spurious effects.
The frequency responses of $P_{\mathrm{on}}(f)$ and $P_{\mathrm{off}}(f)$ 
depend on the relative power of the CW signal to the white noise, 
and also the threshold levels in quantization. 
To correct these effects, 
we need to calibrate the bandpass 
by sensitively adjusting the continuum floor level of $P_{\mathrm{off}}$ to that of $P_{\mathrm{on}}$ 
in the data analysis. 
These values are $a_{\mathrm{on}}$ and $a_{\mathrm{off}}$.

From equations (\ref{eq:oncor}) and (\ref{eq:coroff}), 
the frequency response in a FFX mode is written as 
\begin{eqnarray}
&&P^{\mathrm{F}}_{\mathrm{on}}(f) \nonumber \\ 
&&= a_{\mathrm{on}} \cdot \left[|C(f)|^2 +|N(f)|^2 \right] \cdot H_{1}(f) H_{2}^{\ast}(f) \cdot |H^{\mathrm{F}}_{\mathrm{D}}(f)|^2, 
\label{eq:FFXon}
\end{eqnarray}
while the frequency response without the CW signal can be written as 
\begin{equation}
P^{\mathrm{F}}_{\mathrm{off}}(f) = a_{\mathrm{off}} \cdot |N(f)|^2 \cdot H_{1}(f) H_{2}^{\ast}(f) \cdot |H^{\mathrm{F}}_\mathrm{D}(f)|^2, 
\label{eq:FFXoff}
\end{equation}
and then the CW frequency response including 
the response of the measurement system is derived as 
\begin{eqnarray}
P^{\mathrm{F}}(f) &=& P^{\mathrm{F}}_{\mathrm{on}}-P^{\mathrm{F}}_{\mathrm{off}} \cdot \frac{a_{\mathrm{on}}}{a_{\mathrm{off}}} \nonumber \\
&=& a_{\mathrm{on}} \cdot |C(f)|^2 \cdot H_{1}(f) H_{2}^{\ast}(f) \cdot |H^{\mathrm{F}}_\mathrm{D}(f)|^2. 
\label{eq:FFXpwr}
\end{eqnarray}
Similarly, the frequency response in a FX mode is written as 
\begin{eqnarray}
&&P^{\mathrm{N}}_{\mathrm{on}}(f) \nonumber \\ 
&&= a_{\mathrm{on}} \cdot \left[|C(f)|^2 +|N(f)|^2 \right] \cdot H_{1}(f) H_{2}^{\ast}(f) \cdot |H^{\mathrm{N}}_{\mathrm{D}}(f)|^2, 
\label{eq:FXon}
\end{eqnarray}
while the frequency response without the CW signal can be written as 
\begin{equation}
P^{\mathrm{N}}_{\mathrm{off}}(f) = a_{\mathrm{off}} \cdot |N(f)|^2 \cdot H_{1}(f) H_{2}^{\ast}(f) \cdot |H^{\mathrm{N}}_\mathrm{D}(f)|^2, 
\label{eq:FXoff}
\end{equation}
and then the CW frequency response including 
the response of the measurement system is derived as 
\begin{eqnarray}
P^{\mathrm{N}}(f) &=& P^{\mathrm{N}}_{\mathrm{on}}-P^{\mathrm{N}}_{\mathrm{off}} \cdot \frac{a_{\mathrm{on}}}{a_{\mathrm{off}}} \nonumber \\ 
&=& a_{\mathrm{on}} \cdot |C(f)|^2 \cdot H_{1}(f) H_{2}^{\ast}(f) \cdot |H^{\mathrm{N}}_\mathrm{D}(f)|^2. 
\label{eq:FXpwr}
\end{eqnarray}

From equations (\ref{eq:FFXpwr}) and (\ref{eq:FXpwr}), we can derive 
the frequency response in a FFX mode 
from the correlated spectra obtained in FFX and FX modes as 
\begin{equation}
|H^{\mathrm{F}}_\mathrm{D}(f)|^2 = \frac{P^{\mathrm{F}}(f)}{P^{\mathrm{N}}(f)} \cdot |H^{\mathrm{N}}_\mathrm{D}(f)|^2. 
\end{equation}
Since the frequency response of $H^{\mathrm{N}}_\mathrm{D}(f)$ in the FX mode 
is well known (\cite{TH01}), 
the frequency response in a FFX mode can be derived finally.

Figure \ref{fig:response} shows the frequency response of the FFX correlator 
at a bandwidth of 64 MHz in the range of 2048 to 2112 MHz, 
and that the lower limits of about -33 dB to -40 dB are successfully measured 
with this method. 
The measurement frequency resolution was 31.25 kHz. 
The measurement results show that the effective bandwidth is about 59.28 MHz, 
which was obtained by the passband responses of about $-1$ dB at 2046.40625 and 2106.28125 MHz, 
$-3$ dB at 2045.3125 and 2107.28125 MHz, 
and by a stopband response with the first sidelobe of about $-20$ dB. 
The measurement results are well consistent with 
the theoretical curve in the passband, both bandedges (sharpness), 
and first sidelobe levels. 
In the stopband response except these responses, however, 
it is shown that there are 
differences between theoretical curve and measured results. 
This problem is probably due to 
the non-linear response of 1-bit ADC 
and the precision of the data reduction process in this measurement method. 
The cross-modulation distortion is strongly 
generated in digitizing the CW signals at 1 bit. 
This character complicates the data reduction method, 
and will reduce the measurement precision. 
If the ADCs with 3 bits or more are feasible, 
this problem will be relaxed. 

Finally, the measurement results show that 
the theory of the FFX scheme can be confirmed, and 
the development of the FFX Correlator was successfully realized.

\section{Summary}
There are two basic designs of a digital correlator: the XF-type in which the cross-correlation is calculated 
before Fourier transformation, and the FX-type in which Fourier transformation is 
performed before cross multiplication. 
To improve the FX-type correlator, 
we established a new algorithm for correlation process, 
that is called the FFX scheme. 
The FFX scheme demonstrates 
that the realization of a stopband response 
with first and second sidelobes of $-34$ dB and higher-order sidelobes of $-60$ dB 
is technically feasible. 
The FFX scheme consists of 2-stage Fourier Transform blocks, 
which perform the 1st-stage Fourier Transform as a digital filter, 
and the 2nd-stage Fourier Transform to achieve higher dispersion.
The FFX scheme provides flexibility in the setting of bandwidth
within the sampling frequency.

The input data rate of the developed FFX correlator is 
about 48 Giga bit per second (Gbps)
with 3-bit quantization at the sampling frequency of 
8192 or 4096 Msps, 
which is 8192 Msps x 3 bits x 2 IF or 4096 Msps x 3 bits x 4 IFs. 
We have successfully evaluated the 
feasibilities of the FFX correlator hardware. 
Also, this hardware will be installed and operated as a new spectrometer for ASTE.

We successfully developed the FFX correlator, 
measured its performances, and demonstrated 
the capability of a wide-frequency coverage and high-frequency resolution of 
the correlation systems. 
Our development and measurement results 
will also be useful and helpful 
in designing and developing the next generation correlator.


\section*{Acknowledgments}
The authors would like to acknowledge 
Yoshihiro Chikada 
for his helpful technical discussions. 
The author would like to express gratitude to Brent Carlson 
who provided constructive comments and suggestions on this paper.
This research was partially supported by 
the Ministry of Education, Culture, Sports, Science and Technology, 
Grant-in-Aid for Young Scientists (B), 17740114, 2005.




\begin{thebibliography}{}
\bibitem[Bunton (2000)]{Bunton}
  Bunton,~J, 
  \ 2000, ALMA Memo 342, (Charlottesville: NRAO)
\bibitem[Bunton (2003)]{Bunton}
  Bunton,~J, 
  \ 2003, ALMA Memo 447, (Charlottesville: NRAO)
\bibitem[Carlson (2001)]{EVLA}
   Carlson,~B.
   \ 2001, NRC-EVLA Memo 014
\bibitem[Chikada et al. (1987)]{CY87}
   Chikada,~Y., Ishiguro,~M., Hirabayashi,~H., Morimoto,~M., 
   Morita,~K., Kanzawa,~T., Iwashita,~H., Nakazima,~K., et al. 
   \ 1987, Proc. IEEE, 75, 1203
\bibitem[Escoffier et al. (2007)]{ALMA1}
   Escoffier,~R.~P., Comoretto,~G.
   Webber,~J.~C., Baudry,~A., Broadwell,~C.~M., 
   Greenberg,~J.~H., R. R. Treacy,~R.~R., Cais,~P., et al.
   \ 2007, A$\&$A 462, 801
\bibitem[Ezawa et al. (2004)]{ASTE}
   Ezawa,~H., Kawabe,~R., Kohno,~K., Yamamoto,~S. 
   \ 2004, SPIE 5489, 763
\bibitem[Iguchi et al. (2002)]{IG02} 
   Iguchi,~S., Okuramu,~S.~K., Okiura,~M., Momose,~M., Chikada,~Y. 
   URSI General Assembly (J6:RECENT SCIENTIFIC DEVELOPMENTS), Maastricht, 2002. 
\bibitem[Iguchi et al. (2004)]{IG04} 
   Iguchi,~S., Kurayama,~T., Kawaguchi,~N., $\&$ Kawakami,~K. 
   \ 2005, PASJ 57, 259
\bibitem[Narayanan et al.(2005)]{nar05} 
   Narayanan, D., Groppi, C.~E., Kulesa, C.~A., \& Walker, C.~K.
   \ 2005, \apj, 630, 269 
\bibitem[Okuda et al.(2008)]{oku07} 
   Okuda, T., Iguchi, S. 
   \ 2008, PASJ 60, 315 
\bibitem[Okumura et al.(2001)]{oku01} 
   Okumura, S. K., Chikada, Y., Momose, M., Iguchi, S. 
   \ 2001, ALMA Memo No.350 (Charlottesville: NRAO)
\bibitem[Rabiner and Schafer (1971)]{FSF}
   Rabiner,~L.~R., Schafer,~R.~W. 
   \ 1971, IEEE Trans. Audio Electroacoust., AU-19, 200
\bibitem[Thompson et al. (2001)]{TH01}
   Thompson,~A.~R., Moran,~J.~M., $\&$ Swenson,~G.~W.Jr. 
   \ 2001, Interferometry and Synthesis in Radio Astronomy, 
   2nd Ed., (New York:John Wiley $\&$ Sons), 289
\bibitem[Weinreb (1963)]{XF}
   Weinreb, S. 
   \ 1963, Digital Spectral Analysis Technique and Its Application to Radio Astronomy, 
   (R. L. E., MIT, Cambridge, Mass.), Tech. Rep. No. 412
\end{thebibliography}
\end{document}